\documentclass[prb,twocolumn,showpacs,floatfix]{revtex4-1}
\usepackage[T1]{fontenc}
\usepackage[utf8]{inputenc}
\usepackage{graphicx}
\usepackage{amsmath,amssymb,bm,dcolumn,times}
\usepackage{amsmath}
\usepackage{color}
\usepackage{ulem}

\newcommand{\be}{\begin{equation}}
\newcommand{\ee}{\end{equation}}
\newcommand{\bea}{\begin{eqnarray}}
\newcommand{\eea}{\end{eqnarray}}

\begin{document}

\title{Anderson localization of light in disordered superlattices containing graphene layers}

\author{A. J. Chaves}
\email{andrej6@gmail.com}
\affiliation{Department of Physics and Center of Physics, University of Minho, P-4710-057, Braga, Portugal}

\author{N. M. R. Peres}
\email{peres@fisica.uminho.pt}
\affiliation{Department of Physics and Center of Physics, University of Minho, P-4710-057, Braga, Portugal}

\author{F. A. Pinheiro}
\email{fpinheiro@if.ufrj.br}
\affiliation{Instituto de F\'isica, Universidade Federal do Rio de Janeiro, Rio de Janeiro-RJ, 21941-972, Brazil}
\affiliation{Optoelectronics Research Centre and Centre for Photonic Metamaterials, University of Southampton, 
Highfield, Southampton SO17 1BJ, United Kingdom}

\date{\today}

\begin{abstract}

We theoretically investigate light propagation and Anderson localization in one-dimensional 
disordered superlattices composed of dielectric stacks with graphene sheets in between. Disorder
is introduced either on graphene material parameters ({\it e.g.} Fermi energy) or on the widths of the dielectric stacks. 
We derive an analytic expression for the localization length $\xi$, and compare it to numerical simulations using transfer matrix
technique; a very good agreement is found. We demonstrate that the presence of graphene may strongly attenuate the anomalously 
delocalised Breswter modes, and is at the origin of a periodic dependence of $\xi$ on frequency, in contrast to the usual asymptotic decay,
$\xi \propto \omega^{-2}$.  By 
unveiling the effects of graphene on Anderson localization of light, we pave the way for new applications of graphene-based, disordered photonic 
devices in the THz spectral range.     
\end{abstract}

\pacs{ }

\maketitle

\section{Introduction}

Due to its extraordinary electronic and optical properties, 
graphene has emerged as an alternative material platform for applications in 
photonics and optoelectronics~\cite{Avouris2014,Bludov2013,Zhan2013}. 
A partial but by no means exhaustive list of applications of graphene in 
photonics include high-speed photodetectors~\cite{xia2009}, optical 
modulators~\cite{liu2011}, plasmonic devices~\cite{abajo2014,ju2011,echtermeyer2011}, 
and ultrafast lasers~\cite{sun2010}. In addition, graphene is a promising candidate 
to overcome one of the major existing hurdles to bring optics and electronics together, 
namely the efficient conversion between optical and electronic signals. Indeed, this 
can be facilitated by the fact that graphene enables strong, electric field-tunable 
optical transitions, and resonantly enhances light-mater interactions in sub-wavelength 
volumes. In practice this can be achieved, for instance, by integrating a graphene 
layer into a photonic crystal nanocavity~\cite{engel2012}. The presence of graphene 
also allows for an efficient electro-optical modulation of photonic crystals nanocavities 
by electrostatic gating~\cite{majumdar2013,gan2013}. However, the integration of graphene 
into photonic crystals is naturally prone to unavoidable disorder associated to the fabrication 
process. This constitutes  {\it per se} a motivation to investigate the effects of disorder 
in photonic crystals containing graphene layers which, as far as we know, have not been 
considered in the literature 
so far. In addition to this technological and practical motivation, there is a very fundamental 
one as well, namely to understand the impact of graphene on Anderson localization of light.

The concept of Anderson localization (AL) was originally conceived in the realm of condensed 
matter physics as a disorder driven metal-insulator transition~\cite{anderson1958}. 
Being an interference wave phenomenon, this concept has been extended to light~\cite{segev2013}, 
acoustic waves~\cite{hu2008}, and even Bose-Einstein condensed matter waves~\cite{billy2008}.  
As a result, Anderson localization is today a truly interdisciplinary topic, and important 
contributions have emerged from different areas, ranging from condensed matter, photonics, 
acoustics, atomic physics, and seismology~\cite{ad2009}.  Dimensionality is crucial to AL, 
and in 1D the vast majority of states is exponentially localised on a length scale given 
by the localization length $\xi$, regardless of the disorder strength. In optical systems 
exceptions do exist, and delocalised modes may occur in low-dimensional systems as a result
 of the presence of correlations~\cite{im2009}, necklace modes~\cite{wiersma2005}, or 
metamaterials with negative refraction~\cite{Mogilevtsev2010, Mogilevtsev2011, Asatryan2007}. 
The question of whether these anomalies occur when graphene is integrated into disordered optical 
superlattices  remains an open question.

\begin{figure}[h] 
   \vspace{0.5cm}
   \centering
   \includegraphics[scale=2.5]{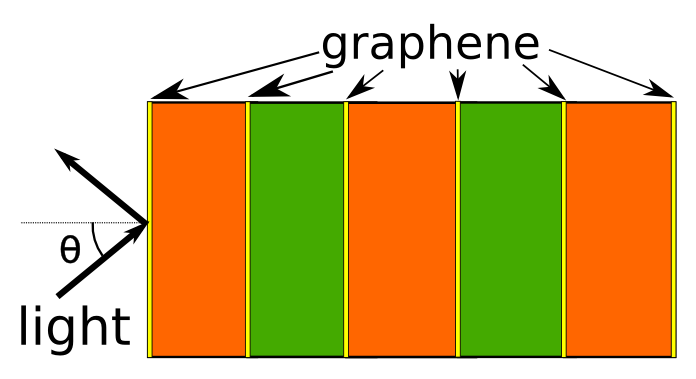}
   \caption{{\it Color on-line.}  Schematic representation of the system.}
   \label{1dhet}
\end{figure}

Bearing in mind both these technological and fundamental motivations, in the present 
paper we undertake an analytical and numerical investigation 
of Anderson localization of 
light in one-dimensional disordered superlattices composed of dielectric stacks with graphene 
layers in between, as depicted in Fig.~\ref{1dhet}.  We consider two possible, realistic 
ways to model disorder: compositional and structural disorder. In the former case disorder 
is introduced in graphene's material parameters, such as the Fermi energy, whereas in the 
latter the dielectric components of the superlattice have random widths. In both cases, we 
derive an analytic expression for the localization length $\xi$, and compare it to numerical 
simulations using a transfer matrix technique; an overall very good agreement is found. 
In the case where the medium impedances match, we find that $\xi$ exhibits an oscillatory
 behaviour as a function of frequency $\omega$, in contrast to the usual asymptotic 
decay $\xi \propto \omega^{-2}$. We demonstrate that graphene may strongly suppress the anomalously
delocalised Brewster modes, as it 
induces additional reflexions at the superlattice interfaces. We also investigate the effects of inter 
and intraband transitions of the 
graphene conductivity on $\xi$, identifying the regimes where Anderson localization and absorption dominates 
light transmission.  

This paper is organised as follows. In Sec. II we present the analytical results, where we derive an expression for 
the
localization length of disordered superlattices containing graphene sheets. 
In Sec. III we present and discuss the numerical simulations, based on transfer matrix technique, which are also compared to the
analytical calculations. Finally, Sec. IV is devoted to the concluding remarks. We also present a number of 
appendices  giving 
the details of the calculations and aiming at making the the text as self-contained as possible. 
To our best knowledge, there are only two published papers\cite{kuzmiak1997,soto2004} dealing with similar problems
to the one we consider in this paper, but in the context a metals, in which case only Drude's conductivity plays a role.
   


\section{Analytical calculation of the localization length} \label{analytic}

Light propagation in a 1D superlattice containing graphene layers (Fig. \ref{1dhet}) is 
modelled by the transfer matrix formalism~\cite{Markos2008}. 
The $M^n=\{m^n_{ij}\}$ transfer matrix   connects 
the fields at the
right of the $n-$th unit cell to those at left according to:
\begin{flalign}
\psi^{n+1}=M^{n}\psi^{n},
\label{eq:Tmatrix}
\end{flalign}
where $\psi^n= \begin{bmatrix}\psi_R^n && \psi_L^n \end{bmatrix}^T$, and $\psi_R^n$ ($\psi_L^n$)
refers to the right (left) propagating field in the $n-$th cell.
For transverse electric (TE) and transvere magnetic (TM) modes,
$\psi$ refers to the electric and magnetic field, respectively.
We consider the particular case where $\det M^{n}=1$, which occurs for systems with
preserved time reversal symmetry \cite{Markos2008}. In this case, one can show that $M^n$ may be written as
\begin{flalign}
M^n=\begin{pmatrix} \cosh\phi_1^n e^{i\phi_2^n} && \sinh\phi_1^n e^{i\phi_3^n} \\ 
\sinh\phi_1^n e^{-i\phi_3^n} &&\cosh\phi_1^ne^{-i\phi_2^n}\end{pmatrix},\label{parmatrix}
\end{flalign}
where $\phi_i^n$ are parameters that depend on the composition of the $n-$th cell.
(from here on we omit
the $n$ dependence in $\phi_i$,  except when strictly necessary to avoid any confusion.)
For periodic systems with preserved time-reversed symmetry, $\phi_i$ are real numbers and all the $M^n$'s are equal. 
We thus write $M^n=M^0$.
 One can write the
photonic dispersion relation~\cite{Markos2008} as 
$\cos\gamma=(m_{11}^0+m_{22}^0)/2 $, where
\be
\cos\gamma=\cosh\phi_1^0\cos\phi_2^0\,.
\ee
Disorder is introduced in the parameters $\phi_i$:
\be
\phi_i=\phi^0_i+\delta\phi_i\,
\ee
where $\delta\phi_i$ describes random fluctuations around the average value, and 
which may have different origins, as it will be detailed later  in  the paper.
For a  periodic system, a transformation
$M_\text{transf}=M_\text{circle}M_\text{real}$ (see appendix \ref{matrix})
exists that maps the variables $\psi^n_{R,L}$ into a new set of variables, denoted by
$Q_n$ and $P_n$, such that
$X^T=[Q^n\,\, P^n]^T=M_\text{transf}[\psi^n_R \,\, \psi^n_L]^T$.
These new variables describe a circle in phase space \cite{Izrailev2009}, with
radius $\sqrt{Q_n^2+P_n^2}$  proportional to the
electric field amplitude.
Applying this transformation to Eq. (\ref{eq:Tmatrix}),
the transformed matrix $M^\prime=M_\text{transf}M^nM_\text{transf}^{-1}$
reads
\be
M^\prime=\begin{pmatrix}E_n&&F_n\\G_n&&H_n\end{pmatrix},
\ee
where:
\begin{eqnarray}
E_n&=&\cosh\phi_1\cos\phi_2-\sinh\phi_1\sin\delta\phi_3,
\label{coef1}
\\
F_n&=&-v^2\left(\cosh\phi_1\sin\phi_2+\sinh\phi_1\cos\delta\phi_3\right),
\label{coef2}
\\
G_n&=&v^{-2}\left(\cosh\phi_1\sin\phi_2-\sinh\phi_1\cos\delta\phi_3\right),
\label{coef3}
\\
H_n&=&\cosh\phi_1\cos\phi_2+\sinh\phi_1\sin\delta\phi_3.
\label{coef4}
\end{eqnarray}
with $v$ and $\tau$ defined in appendix \ref{matrix}. When $\phi_i=\phi_i^0$, we have $\delta\phi_3=0$ and
Eqs. (\ref{coef1})-(\ref{coef4}) lead to $E_n=H_n=\cos\gamma$ and $F_n=-G_n=\sin\gamma$.
When weak disorder is introduced, the trajectory  of the points $(Q_n,P_n)$ results
in a perturbation of the circle. The recurrence equations defined by $X^{n+1}=M^\prime X^n$
are similar to a Hamiltonian map of the classical harmonic oscillator subjected to
a parametric impulsive force\cite{Izrailev1995}, where $Q_n$ and $P_n$ are the coordinate and conjugated
moments, respectively, and $\gamma$  is the phase between successive kicks.

The presence of disorder introduces a key length scale, the localization length $\xi$. 
In 1D  electronic systems  all eigenmodes are exponentially localised, although some exceptions do exist in 
 the realm of 
optical systems~\cite{Mogilevtsev2010, Mogilevtsev2011, Asatryan2007,wiersma2005} (see Introduction).  The length  $\xi$
 characterises the exponential decay of the eigenfunctions and is defined in terms of 
the reciprocal of the Lyapunov exponent $\lambda$. In 1D $\lambda$ can be written as 
\cite{Markos2008,Izrailev2009}:
\be
\lambda = \frac{1}{2} \left\langle \ln\left|\frac{\psi^{n+1}_R}
{\psi^n_R}\right|^2\right\rangle\,.
\label{lamb}
\ee
In Eq. (\ref{lamb}) 
the brackets  denote  averaging over both ensembles and the  system unit cells,
 while the usual definition of the localization length 
considers only averages over ensembles~\cite{Markos2008}. The two definitions are equivalent.
 The relation between $\lambda$ and $\xi$  is:
\be
\text{Re} \lambda=\frac{d}{\xi},
\ee
where $d$ is the mean length of the unit cell. 
The advantage of the approach based on the parameters $P_n$ and $Q_n$
is that we can use polar (or action-angles) coordinates:
\bea
P_n=R_n\sin\Theta_n,\nonumber\\
Q_n=R_n\cos\Theta_n\,.
\eea
 
Without disorder, $R_n$ is a constant and $\Theta_n$ increases by minus the Bloch
phase,  $-\gamma$, as we move from unit cell to unit cell. 
 With disorder, the
radius $R_n$ changes in every step, with $R_{n+1}$ a function of
$R_n$, $\Theta_n$, and of the matrix elements of $M^n$. The angle $\Theta_{n+1}$ only depends on
$\Theta_n$ and $M^n$. 
For weak disorder a recurrence equation (\ref{eqTheta}) exists that, 
in the continuum limit, becomes a stochastic \^Ito equation which has a corresponding Fokker-Planck 
equation \cite{Izrailev1998} . In this case, the
first approximation for the density probability function of $\Theta_n$ is uniform in the 
interval $[0,2\pi]$ for $\gamma\ne0,\pi/2,\pi$.

Writing  Eq. (\ref{lamb}) in terms of $R$ and $\Theta$, and averaging over
 $\Theta$ with uniform density probability, we obtain, up to second order in $\delta \phi_i$: 
\be
\lambda=\frac{1}{2}\left\langle Y_1+Y_2\cos2\Theta_n+Y_3\sin2\Theta_n-\frac{1}{4}Y_2^2-\frac{1}{4}Y_3^2 \right\rangle, \label{Lyp}
\ee
where $Y_i$, with $i=1,2,3$ are defined in Appendix \ref{formalism} and depend 
on the matrix elements $M^n$.

In the following sections we will study the propagation of light through a disordered structure of alternating
graphene sheets and dielectric layers. In this case each propagation matrix $M^n$ is 
 determined  by the widths $z_i$, the incidence angle $\theta_i$,
the dielectric material parameters $\mu_i$ and $\varepsilon_i$,
and the graphene conductivity $\sigma$.  
In the present work we focus on the cases where disorder is present in the widths of the stacks (structural disorder)
and on graphene
conductivities (compositional disorder) . Both are realistic situations that may occur in the fabrication
of these structures. 

For the type of structural disorder studied here  the width of each layer $i$ of 
the $n^{\rm{th.}}$ cell is a random variable
\be
z_i(n)=z_i^0+\zeta_i(n), \label{dis1}
\ee
 where $\zeta_i$ are uncorrelated random variables with zero mean and
 mean standard deviation  $\sigma_i$: $\langle \left[\zeta_i(n)\right]^2\rangle=\sigma_i^2$; $z_i^0$ is the
mean width of the $i$ slab (the standard deviation  $\sigma_i$ should not be confused with graphene's conductivity $\sigma$).

In the  case of  compositional disorder, the Fermi energy $E_F$ is a random variable in
each layer $n$ 
\be
E_F(n)/\hbar=\omega_F(n)=\omega_F^0+\zeta_F(n), \label{dis2}
\ee
with $\zeta_F$ a random variable with zero mean and $\langle \zeta_F^2\rangle=\sigma_F^2$. This determines how the
graphene conductivity, given in  Appendix \ref{grcond}, is affected by disorder.

In the next section
we derive analytical expressions for $\lambda$ (Eq. \ref{Lyp}) in different  regimes.
To this end, we need to map $\phi_i$ in the
system variables, calculate the differentials $\delta_i$, and use the results
given in  Appendix \ref{formalism}.

\subsection{Unit cell made of two different dielectric materials and a graphene sheet at the interface}

We consider a  disordered superlattice composed of dielectric bilayers with a graphene sheet in between. The  
transfer matrix for the $n$ unit cell is given by $M^n=\{m_{jl}^n\}$ and is explicitly derived in Appendix \ref{mFull}. 

To proceed with the calculation of the Lyapunov exponent  it is necessary
to map the system parameters of the transfer matrix (\ref{mFull}) into
the parametric matrix (\ref{parmatrix}).
There is not a unique way of doing
this, but in what follows we  make the simplest choice.

\subsubsection{Disordered photonic super lattice without graphene} \label{grapheneless}

To model a disordered photonic super-lattice without the graphene layer we put $f=0$ in Eq. (\ref{mtransfer}) and
map $\phi_i$ into the system parameters $\alpha_1,\alpha_2,\chi,\Delta$ (defined in Appendix
\ref{phocry}):
\bea
\sinh\phi_1&=&\Delta^x\sin\alpha_2,\nonumber\\
\phi_2&=&\alpha_1+\arctan\left(\chi^x\tan\alpha_2\right),\nonumber\\
\phi_3&=&\alpha_1+\pi/2,
\eea
where $x=$TE,TM. According to this mapping we can replace $\phi_i$ in the expressions 
for $Y_i$ in Appendix \ref{formalism} and calculate the
differentials $\delta \phi_i$ using Eq. (\ref{dis1}) with $\zeta_i\ll z_i^0$.  This will enable us to compute the Lyapunov
exponent, given by Eq. (\ref{Lyp}); the final result is
\be
\lambda=\frac{\Delta^2}{2\sin^2\gamma}\left(\sin^2\alpha_2\, k_1^2\,\sigma_1^2+\sin^2\alpha_1\, k_2^2\,\sigma_2^2\right), \label{eqIM}
\ee
 which agrees with the result of Ref. [\onlinecite{Izrailev2009}] for uncorrelated disorder. 
 The described procedure is repeated to calculate the Lyapunov exponents in the 
 next sections. 

\subsubsection{Disordered superlattice containing graphene layers}

 The presence of graphene 
at the interface between the dielectrics results in a discontinuity in the tangential
component of the 
magnetic field. The role of graphene  on the optical properties of the superlattice  increases as the value of the
dimensionless parameter $\beta_i^x f$ increases, with $f=\sigma c \mu_0/2$
and $\beta_i^x$ given in Appendix \ref{phocry}. We are interested in the lossless regime  in which the 
Bloch phase $\gamma$, given by Eq. (\ref{disrel}), is real. This regime  
sets up when (i) $\sigma$ (and therefore $f$) is
a pure complex number and $\theta_i$, with $i=1,2$, is a pure real number; or (ii) $\sigma$ is a pure real number so that
evanescent propagation occurs in one of the layers.

In the first case, we define $B^x=i\tilde{B}^x$ 
(see Appendix \ref{phocry}), where $\tilde{B}$ is real, and we
map the parameters $\phi_i$ in:
\bea
\sinh\phi_1&=&-\tilde{B}^x\cos\alpha_2+(\Delta-D^x)\sin\alpha_2,\nonumber\\
\phi_2&=&\alpha_1+\arg\left[A^x_+\cos\alpha_2+i(\chi+C^x_+)\sin\alpha_2\right],\nonumber\\
\phi_3&=&\alpha_1+\pi/2\,.
\eea
Following the procedure of Sec. \ref{grapheneless}, the Lyapunov exponent is given by:
\be
\lambda=\frac{1}{2\sin^2\gamma}\left(K_2^2k_1^2\sigma_1^2+K_1^2k_2^2\sigma_2^2\right), \label{lygr}
\ee
where:
\be
K_1= -2\tilde{f}\lambda^x\beta_2^x\cos\alpha_1+\left[-\Delta+2\tilde{f}^2\lambda^x\beta_1^x\beta_2^x\right]\sin\alpha_1, \label{ki}
\ee
and $f=i\tilde{f}$, $K_2$ is obtained by interchanging $1\leftrightarrow2$ and $\Delta\rightarrow-\Delta$.
 Notice that if one plugs Eq. (\ref{ki}) with $f=0$ into Eq. (\ref{lygr}), 
 Eq. (\ref{eqIM}) is obtained, as it should be. 

\subsection{Unit cell made of one dielectric material and a graphene sheet at the interface}

For systems composed of bilayers of the same dielectric material with a graphene sheet in between,  it is much easier to calculate
the transfer matrix, which is given in Eq. (\ref{1dg}). In this case, the $\phi_i$ parameters read
\bea
\sinh\phi_1&=&-\lambda^x \beta^x \tilde{f},\nonumber\\
\phi_2&=&\alpha,\nonumber\\
\phi_3&=&\alpha+\pi/2, \label{phi1d}
\eea
 Using Eq. (\ref{Lyp}) and the results of the Appendix \ref{formalism} we calculate the Lyapunov exponent
for structures containing both random graphene conductivities
(compositional disorder) and random widths (structural disorder), as detailed in the following. 

\subsubsection{Compositional disorder}

Using the same procedure of  subsection \ref{grapheneless}, we obtain the Lyapunov exponent:
\be
\lambda=\frac{1}{2} \left(\frac{\sin 2\alpha}{\sin 2\gamma} {\beta^x} \frac{\pi\alpha_c}{2} \sigma_{g}\right)^2, \label{lyco}
\ee
where $\alpha_c$ is the fine structure constant and $\sigma_g$ is the
mean standard deviation of the normalized graphene conductivity
\be
\sigma_g^2= \frac{\langle \sigma^2\rangle-\langle\sigma\rangle^2}{\sigma_0^2}. 
\ee
\subsubsection{Structural disorder}

For structural disorder where the stacks'  widths are given by Eq. (\ref{dis1}), the Lyapunov exponent reads
\be
\lambda=\frac{ \tilde{f}^2{\beta^x}^2k^2\sigma^2}{2\sin^2\gamma}. \label{str1D}
\ee

This concludes the analytical part of our work, which 
shall be compared to numerical simulations  in the following section.
 
\section{Numerical Simulations: Results and Discussions}

\subsection{Simulation procedure}

The numerical calculations are based on the transfer matrix method; the total 
transfer matrix for light propagating in a $N$-layered system is
\be
M=\Pi_{n=1}^N M^n. \label{tm}
\ee
where the elements of $M^n$ are given by Eq. (\ref{mFull}).  Transmission 
is calculated by applying the boundary condition related to the fact that  there is 
no incoming wave from the left:
\be
T=\frac{1}{|m_{22}|^2},
\ee
and the localization length $\xi$ is calculated by:
\be
\frac{L}{\xi}=-\frac{1}{2}\langle\ln T\rangle, \label{xi}
\ee
where $L=Nd$ and 
 $N$ is the total number of unit cells with mean width $d$.
The length $L$ is chosen to be large enough to ensure  the numerical 
calculation of the localization length converges.
 In   the  numerical procedure we first generate random variables
 $\zeta_i$ (or $\zeta_F$) [see Eqs. (\ref{dis1}) and (\ref{dis2})]  from  a uniform
 distribution, and then calculate the transfer matrix using Eq. (\ref{tm}). 
 With the help of the results  introduced in Appendix \ref{phocry}, we 
obtain the localization length using Eq. (\ref{xi}). The procedure is repeated over 
$n_\text{samples}$ and the mean value of the localization length is calculated. 
 We have verified that,  for
a sufficiently large  $N$, the value of $\xi$  calculated for a single disorder 
realisation  coincides with its average over many disorder realisations for 
smaller systems; in other words, we have verified that $\xi$ is a self-averaging quantity. 
Further details of the transfer matrix method are given in Appendix \ref{phocry}.

\subsection{Results}

Light transmission depends on the graphene conductivity $\sigma$ and on the medium impedances, 
defined as ( see Appendix \ref{phocry}):
\be
Z^\text{TE}_i=\frac{\sqrt{\mu_i\varepsilon_i}}{\mu_i}\cos\theta_i, 
Z^\text{TM}_i=\frac{\sqrt{\mu_i\varepsilon_i}}{\varepsilon_i}\cos\theta_i. \label{eqimp}
\ee

We shall focus in the lossless regime with $\Im m\cos\gamma=0$ and $\Re e \cos\gamma\le1$.
From Eq.~(\ref{disrel}), this regime occurs whenever $f$ (and consequently $\sigma$) is a
pure complex number or for $\Im \sigma = 0$,  in which case one of the slabs  supports  a
evanescent mode. When the Drude term dominates, the imaginary part of the conductivity is positive (see Appendix \ref{grcond}). 
For frequencies slightly below $2\omega_F$, the inter-band term
dominates and the imaginary part of the conductivity is
negative (see Appendix \ref{grcond}). When the frequency becomes larger than $2\omega_F$, 
the imaginary part goes to zero and the real part tends to $\sigma_0=e^2/4h$.

In the following numerical calculations, random variables have a uniform distribution 
with $\zeta_x\in[-\Upsilon_x/2,\Upsilon_x/2]$, with 
$x=1,2$ for structural disorder and $x=F$ for compositional disorder.

\subsection{Drude regime when: $\Re e\sigma\approx0$, $\Im m\sigma>0$}

When $\omega_F\Gamma\ll\omega^2\ll\omega_F^2$ (where $\Gamma$ is the broadening entering in the 
conductivity), graphene
conductivity can be approximated by (see  Appendix \ref{gcond}):
\be
\sigma=i\sigma_0\frac{4}{\pi}\frac{\omega_F}{\omega}. \label{drudeh}
\ee
For $E_F\approx0.3$ eV (a typical value for the 
graphene Fermi energy), the range of frequencies corresponds
to the infrared spectral regime. 
 In the following  we focus in three  regimes: impedance matching in the double layered system 
[$Z_1=Z_2$, in Eq. (\ref{eqimp})]
with structural disorder, compositional disorder in one layered system,
and the attenuated field regime (ATR) with structural disorder.

\subsubsection{Impedance matching in two-layered system with structural disorder}

Using the Snell-Descartes law, Eq. (\ref{snell}), and the impedances in Appendix \ref{phocry}, 
 one can verify  that for
materials without magnetic response ($\mu_1=\mu_2=1$), there is no TE mode that
allows the impedance matching. In the TM mode  the impedance matching occurs
when the angle of incidence in layer $1$ obeys the relation 
$\sin^2\theta_1=\varepsilon_2/(\varepsilon_1+\varepsilon_2)$, 
for $\varepsilon_1\ne\varepsilon_2$.

\begin{figure}[h] 
   \includegraphics[scale=0.18]{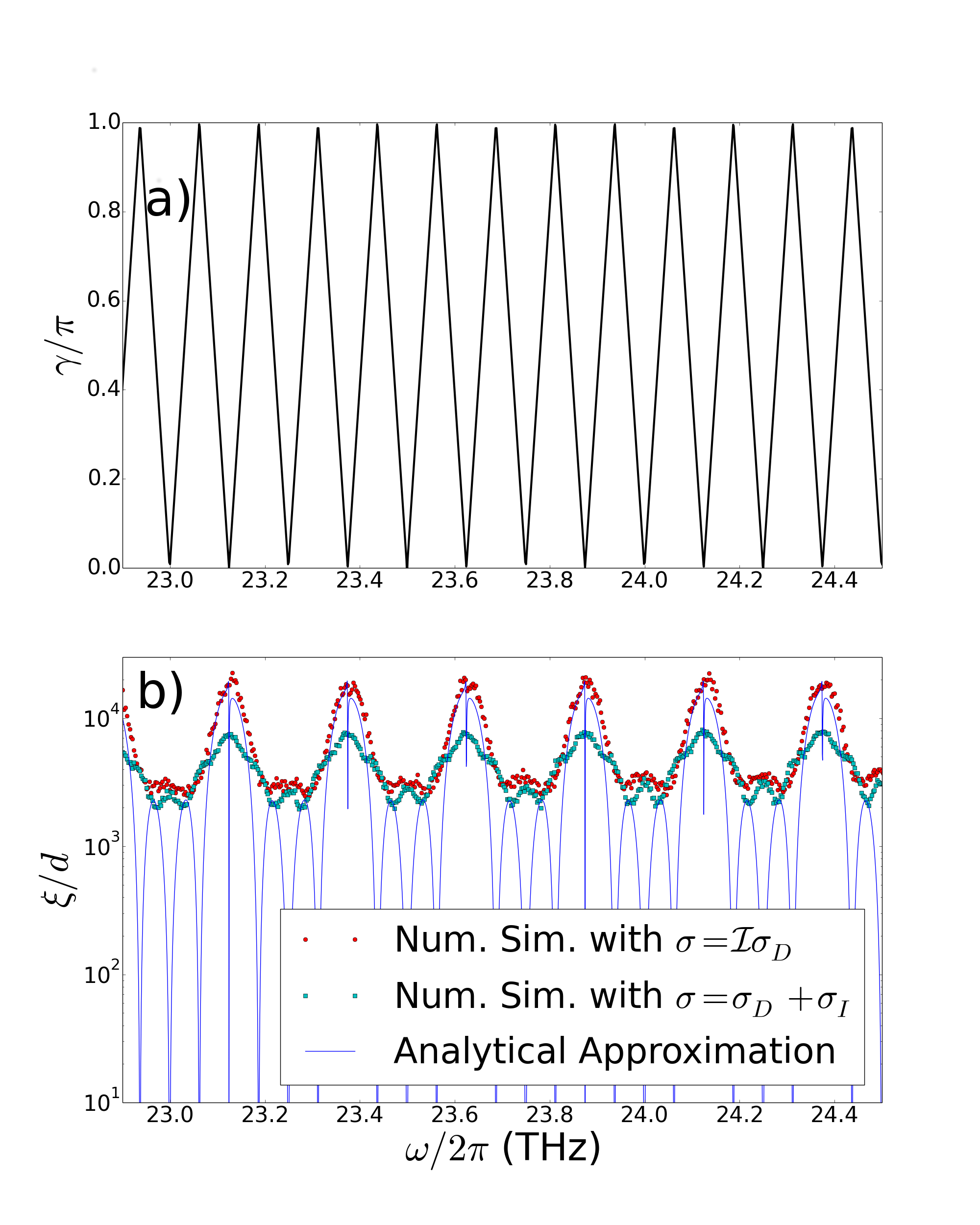}
   \caption{{\it Color on-line.} (a) Dispersion relation in the impedance matching regime and TM mode. (b) Localization length as a function of frequency with $\Upsilon_i=5\mu$m, $z_i^0=1.2$mm, $E_F=0.2$ eV, $\Gamma=260. \mu$eV, $N=5000$, $n_\text{samples}=100$, $\varepsilon_1=\mu_1=\mu_2=1$, 
    $\varepsilon_2=3$, $\theta=\pi/3$.The solid line in (b) is the analytical result, whereas the dots correspond to two different numerical 
simulations for different regimes of the optical conductivity of graphene: (i) $\sigma=\Im m\sigma_D$
(red points) and (ii) $\sigma=\sigma_D+\sigma_I$ (blue points).}
   \label{impmat}
\end{figure}

When $Z_i=Z$, $\beta_i=\beta$,   it follows  from Eqs. (\ref{lygr}) and Eq. (\ref{ki}) that:
\be
\lambda=2\left(\frac{4\tilde{f}\beta^x\omega_F }{\pi c\sin\gamma}\right)^2\sum_{i=1}^2\, \varepsilon_i\mu_i\cos^2\alpha_i\cos^2\theta_i\sigma_i^2,
\ee
where we neglected the term $\tilde{f}^2$  in comparison to  $\tilde{f}$ 
(which in the Drude regime is always valid for a sufficient
large $\omega$).
 In this case, in Fig. \ref{impmat}b the localization length $\xi$ is calculated, both analytically and 
numerically, as a function of frequency. The dispersion relation is also shown in Fig. \ref{impmat}a. It is 
important to point out that the agreement between the analytical and numerical calculations is very good, 
except when $\gamma$ approach $0$ or $\pi$. This is due to the fact that, in the analytical derivation of the Lyapunov 
exponent, the recurrence equation (\ref{eqTheta}) is ill defined at these points, so that the distribution of 
random variables is not uniform.
Remarkably, Fig. \ref{impmat}b reveals that in the impedance matching regime, $\xi$ does not follow the well-known 
asymptotic power law $\omega^{-2}$ behaviour for low frequencies. Rather, $\xi$ exhibits a periodic dependence on $\omega$
for low frequencies, a result that is intrinsically related to the graphene conductivity properties. Indeed, it can 
be explained by the fact that  
the linear increase of the wavenumber with frequency is cancelled by the simultaneous decrease of graphene's
conductivity (Drude term, see Eq. \ref{drudeh}), which scales with $1/\omega$.  The periodicity in 
$\xi$ follows from the periodicity in the dispersion relation,  shown in Fig. \ref{impmat}a. 
For the lossy and Drude regimes, $\xi$ approaches the same value as the frequency increases, and the real 
part of the Drude conductivity goes to zero.  

\begin{figure}[h] 
   \includegraphics[scale=0.18]{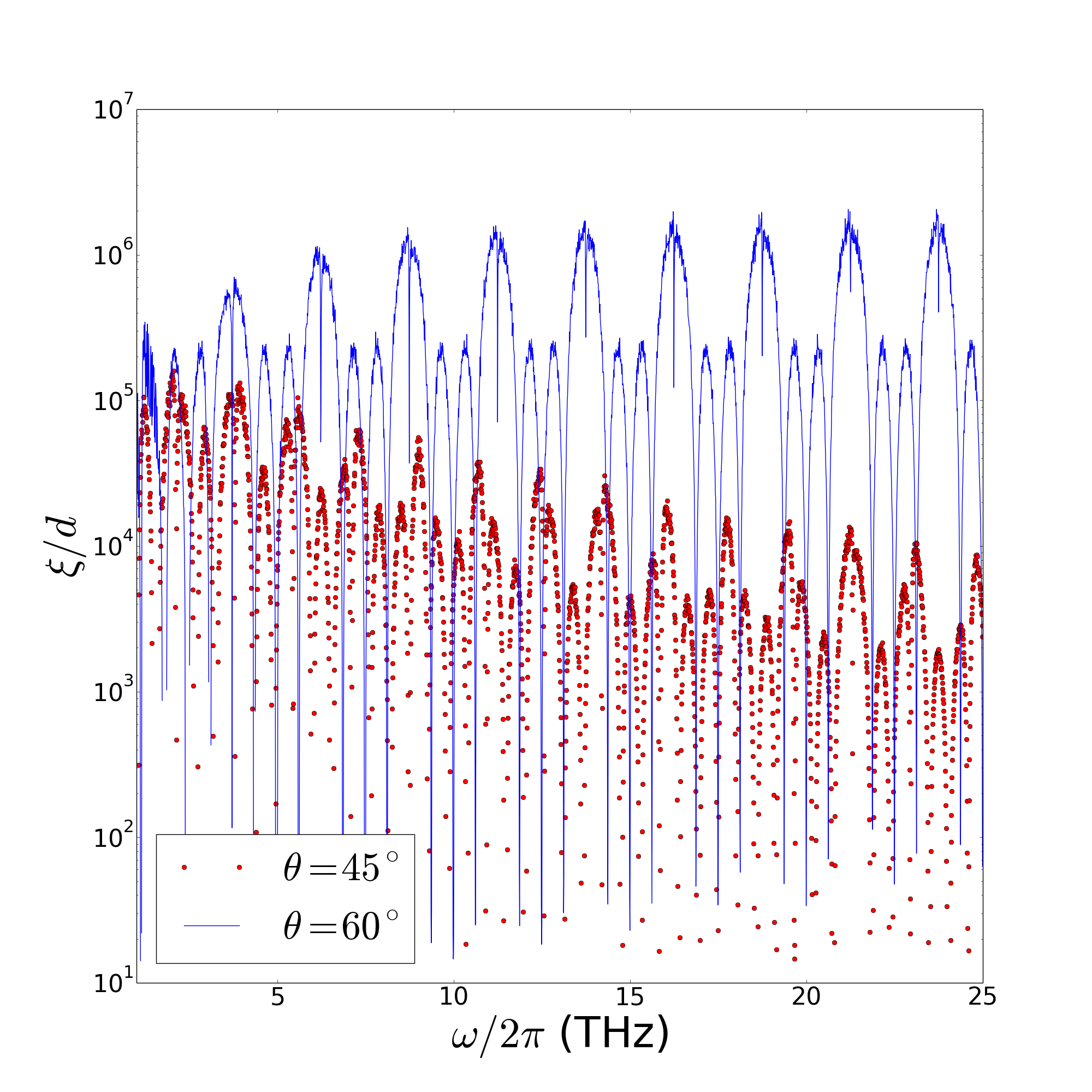}
   \caption{{\it Color on-line.} Localization length as a function of frequency in the impedance matched regime for two values of incidence angle: $\theta=60^\circ$ (blue line), $\theta=45^\circ$ (red circles).  $\Upsilon_i=0.5\mu$m, $z_i^0=120\mu$m, $E_F=0.2$ eV, $\Gamma=0. \mu$eV, $N=5000$, $n_\text{samples}=100$.}
   \label{theta}
\end{figure}

\begin{figure}[h] 
   \includegraphics[scale=0.18]{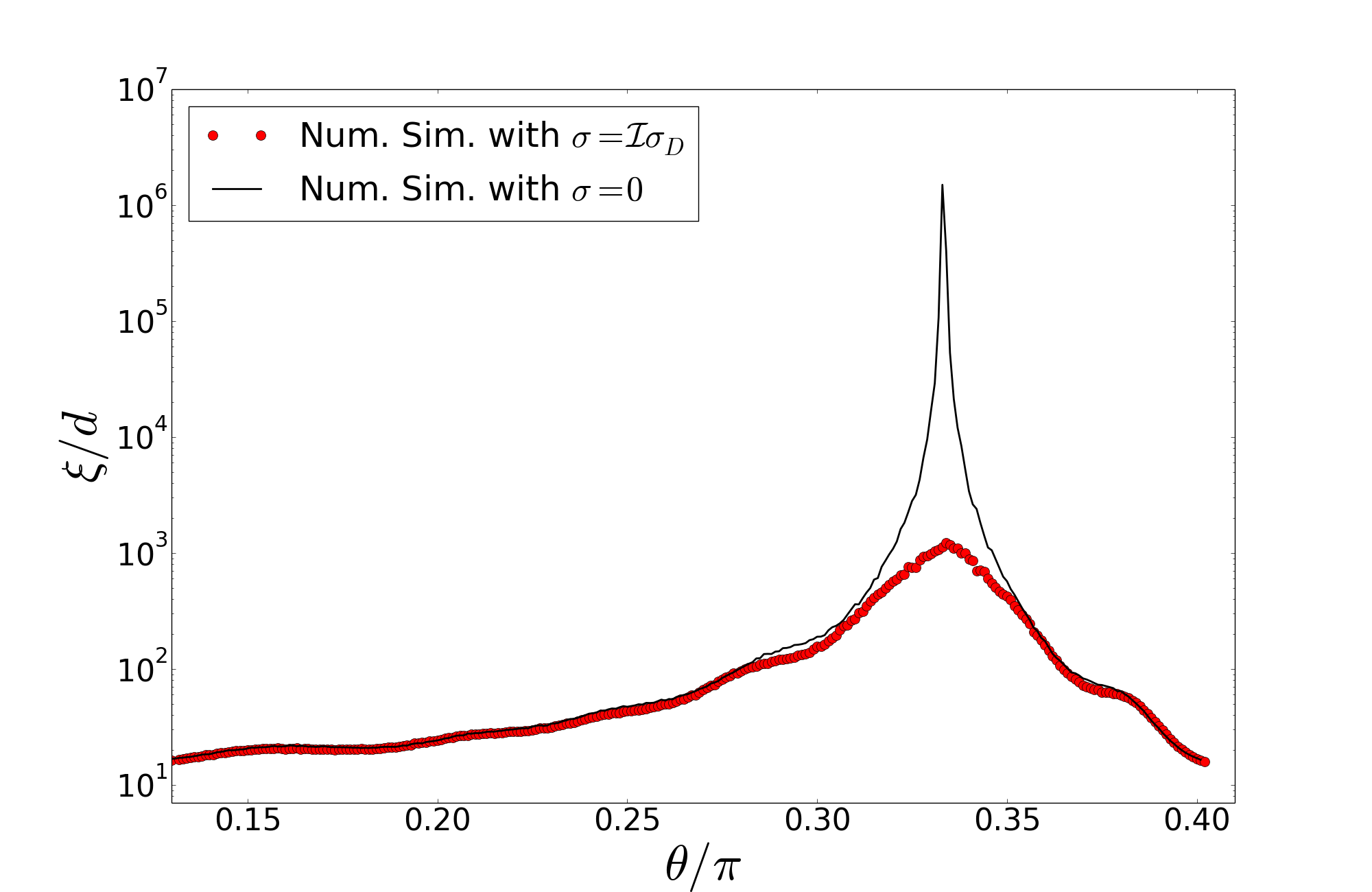}
   \caption{{\it Color on-line.} Localization length as a function of incidence angle in the impedance matched regime at the vicinities of a Brewster mode.  $\Upsilon_i=50\mu$m, $z_i^0=120\mu$m, $N=5000$, $n_\text{samples}=100$. The black solid line corresponds to the grapheneless case; red circles correspond to the case where graphene is present in the superlattice ($E_F=0.2$ eV and $\Gamma=0\, $eV.)}
   \label{theta2}
\end{figure}

 Figure \ref{theta} shows $\xi$ as a function of frequency for two different values 
of the incidence angle $\theta$. It reveals that the presence of graphene layers
has also an important effect in the so-called Brewster modes in disordered systems. 
In 1D disordered optical systems, the so-called Brewster modes occur at some specific frequencies
and incident angles for which $\xi$ reaches anomalously high values, larger than the 
system size~\cite{Sipe1988,Mogilevtsev2010}. For non-magnetic ($\mu_1=\mu_2=1$) superlattices made
of positive refractive-index media, these anomalously delocalised modes arise from the suppression of 
reflexion at the interfaces of a 1D disordered system illuminated by a TM 
incident wave~\cite{Sipe1988,Mogilevtsev2010}. As a result, the system becomes
fully transparent. The presence of graphene induces
additional reflections at each interface of the superlattice, resulting in an attenuation of this Brewster mode, as it 
can be seen from fig. \ref{theta2}.

\subsubsection{ATR regime in one-layered system}

The plasmon-polariton mode in graphene can be excited for example, by a prism
in the Otto configuration [\onlinecite{bludov2010}]. This is the regime we will explore in this 
section. We consider a periodic array of graphene/air unit cells (medium $2$)   in 
between a dielectric (medium $1$). In this case the total transfer matrix $M$
is obtained considering the boundaries between the prism and the superlattice:
\be
M=M_{1\rightarrow 2} \,\prod_j(M_j)\, M_{2\rightarrow1},
\ee
where $M_{1\rightarrow2}$ refers to the transfer matrix describing light propagation from the medium $1$ (dielectric) to 
medium $2$ (air);  
$M_{2\rightarrow1}$  refers to the reverse propagation. $M_j$ is the transfer matrix of the
 unit cell air/graphene with random widths (medium $2$).

From Eq. (\ref{cosg1d}) one can see that for the evanescent mode 
$\alpha$ is a pure complex number and the first term in
the right hand side becomes a hyperbolic cosine, which is greater than $1$ for
any $\alpha$.  As a  result,  the Bloch phase is real only if the second term
in the right  hand side of \ref{cosg1d} is negative. This situation occurs for pure 
positive complex $f$;  in this case  $\beta$ is also a pure positive complex number,   
which is only possible in the TM mode [see Eqs. (\ref{compimp}) and (\ref{betas})].

For an incidence angle $\theta_1$ above the critical angle 
for total reflection at the interface $1/2$, a plasmon-polariton
can be excited, allowing for frustrated total internal reflection. 
In this case light propagation occurs due to the presence of periodic graphene sheets. 
The effective impedance in the  medium $2$ depends on the properties of the layer $1$ as:
\be
Z_2^\text{TE}=i\frac{\kappa}{\mu_2},\,\,\,\, Z_2^\text{TM}=i\frac{\kappa}{\varepsilon_2}, \label{compimp}
\ee
where
\be
\kappa=\sqrt{\varepsilon_1\mu_1\sin^2\theta_1-\varepsilon_2\mu_2}.
\ee

\begin{figure}[h] 
   \centering
   \includegraphics[scale=0.18]{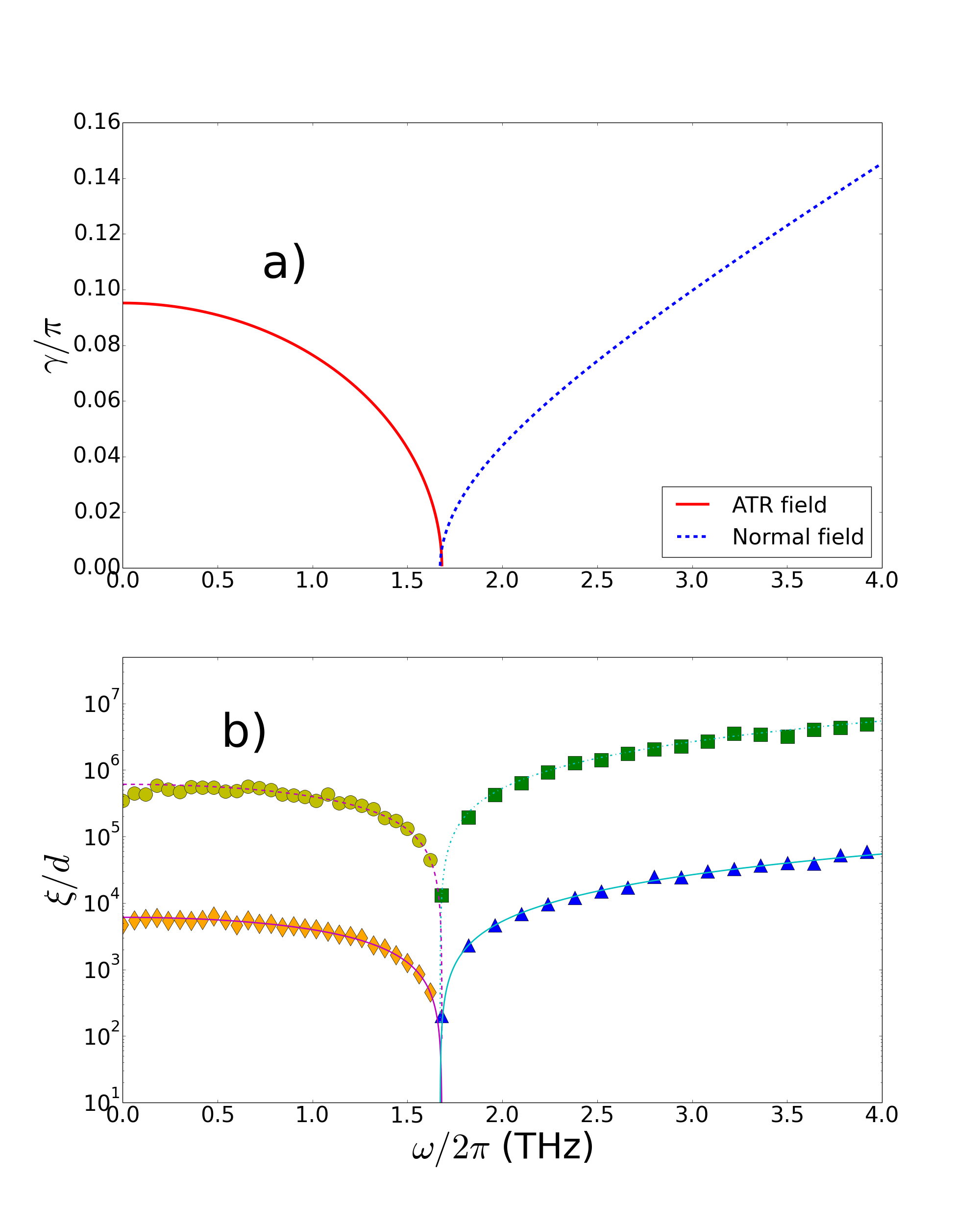}
   \caption{{\it Color on-line.} (a) Dispersion relation for the ATR regime (red) 
   and for normal field (blue). (b) Localization length as a function of frequency for $z^0=12 \mu$m, $E_F=0.1$ eV, $\theta_1=\pi/3$, 
$\varepsilon_1=2$,$\mu_1=\mu_2=\varepsilon_1=1$, $N=50000$, $n_\text{samples}=1$. 
The yellow circles (green squares) and orange diamonds (blue triangles) refer to the 
ATR (normal) field  with  $\Upsilon=0.5 \mu$m and $\Upsilon=5\mu$m, respectively. 
The cyan and
purple lines refer to the analytical approximation.}
   \label{atr}
\end{figure}

In Fig. \ref{atr}  the localization length is calculated in the ATR regime 
 using  both numerical and analytical methods; the agreement is excellent. 
In the Drude regime $\xi$ is inversely proportional to the Fermi energy. 
Also shown is the localization length when  the dielectric necessary to excite the ATR field is removed;
we call this situation the normal field. 
The ATR field is characterized by exponentials with argument $\pm\omega\kappa z/c$.
 When the frequency
increases and the length $c/\kappa\omega$ becomes smaller than the
width $z$ of the dielectric slab (air in this case)  light propagation comes to a halt, as the plasmon-polariton
localized in a graphene layer cannot excite the adjacent  layer. We can see
that the ATR for the parameters of Fig. \ref{atr} fills the band gap of the normal field. Also the increase
of disorder implies  in the decrease of $\xi$, as expected. 

 Notice that  ignoring the interband term and making $E_F=0$ is equivalent to
remove the graphene sheets, therefore making disorder in random widths of air  
meaningless. Hence  
the localization length diverges, as can be seen in Eq. (\ref{str1D}), 
where $\sigma\rightarrow0$ implies in a   vanishing  Lyapunov exponent.

\subsubsection{One layer system with compositional disorder}

In the compositional disorder regime and for the one layered  system, $\xi$ decreases as 
$\beta$ increases. For the TE mode, $\beta$ can only  be  greater than $1$
for materials with magnetic response, $\mu>1$. For the TM mode, $\beta$ is proportional 
to the dielectric constant and to $\cos\theta$, thus for grazing incidence,
the system becomes fully opaque.

In the Drude regime the asymptotic behaviour of the localization
length  goes as $\omega^2$. This can be understood as follows: as the frequency increases
the graphene conductivity decreases as $\omega^{-1}$ and thus the influence 
of the graphene layer disappears.

\begin{figure}[h] 
   \centering
   \includegraphics[scale=0.18]{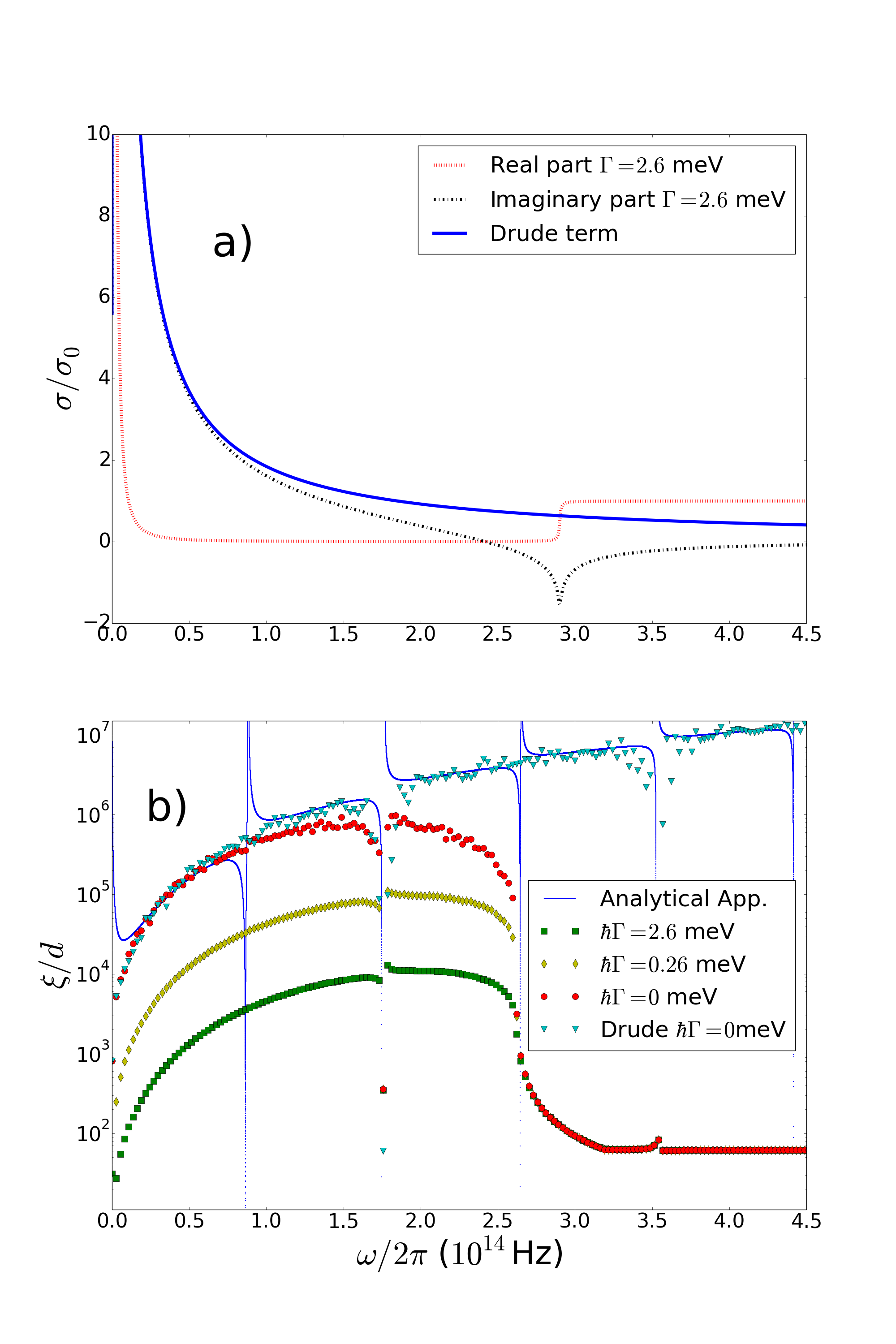}
   \caption{{\it Color on-line.} (a) Real and imaginary parts of the graphene optical conductivity in the compositional disordered case,  $\sigma=\sigma_D+\sigma_I$, 
and the Drude conductivity $\sigma_D$ when $\Gamma=0$. (b)  Localization length as a function of frequency with $z^0=1.2 \mu$m, $\theta=\pi/4$, 
$\varepsilon=\mu=1$, $E_F^0=0.6$ eV, $\Upsilon_F=0.12$ eV, $N=5000$ and 
increasing relaxation rate $\Gamma$.
$\omega_F\approx3\times10^{14}$ Hz. The blue triangles (and solid blue line) 
refer to a calculation where
only the Drude conductivity with $\Gamma=0$ is used. The other data sets refer to
the use of the full optical conductivity of graphene with different $\Gamma$ values.}
   \label{fermi}
\end{figure}

The effect of compositional disorder is  shown in Fig. \ref{fermi}, where the
Fermi energy  is randomly distributed around the mean value $E_F^0=0.6$ eV. 
$\xi$ is inversely proportional to the mean standard deviation of the Fermi energy. 
We study the effect of increasing absorption in graphene layers, which depends on the real part of the 
conductivity and is 
proportional to the relaxation rate $\Gamma$.  The length  $\xi$ decays rapidly 
when the frequency reaches $2\omega_F$, and interband transitions start  to occur, 
an effect that  may be  related  either  to absorption or to Anderson localization. 
The numerical calculation is performed with the full graphene conductivity (Drude plus interband) and  
then compared to the case where only the Drude term is present. The analytical approximation
is calculated with the Drude term only, and 
agrees  very well  with the numerical simulation except at the band edges $\gamma=0,\pi/2,\pi$.
 As already discussed, this disagreement is related to the fact that the probability distribution of 
$\Theta_n$ is not uniform for these values of $\gamma$. 
The analytical approximation has a peak at  $\gamma=\pi/2$ (see denominator of Eq. \ref{lyco}).
The numerical calculations show that near the band gap 
($\gamma=0,\pi$ see Eq. \ref{lyco}) $\xi$ goes to zero, and the peak at $\gamma=\pi/2$ does not occur.

\subsection{Complex interband regime when: $\Re e\sigma\approx0$, $\Im m\sigma<0$}

When $\omega\lesssim2\omega_F$, the imaginary part of the optical conductivity of graphene
becomes negative and can be approximated by 
\be
\sigma=i\sigma_I^{\prime\prime}+i\sigma_0\frac{4}{\pi}\frac{\omega_F}{\omega},
\ee
where $\sigma_I^{\prime\prime}$ is given by Eq. (\ref{interimag}). In this case the imaginary part of $\sigma$ becomes negative,
and the ratio between the imaginary and real parts of $\sigma$ becomes
lower than in the Drude regime for typical values of $\Gamma$ and $E_F$. Therefore, in 
this case the exponential decay of transmission is essentially due to absorption rather than to
Anderson localization. Therefore, in this case, our approach for studying Anderson localization using the localization length is inadequate.
It is worth commenting that experimentally it is possible to distinguish between absorption
and Anderson localization by investigating the variance of the 
normalized total transmission, as proposed in Ref. [\onlinecite{Chabanov2000}]. 
For a one-layered  system in the ATR regime with transfer matrix given by Eq. (\ref{1dg}), the 
change in the sign of $f$ has qualitatively the same effect in the dispersion relation (\ref{cosg1d})  of interchanging  TE and TM modes, which  changes  the sign of $\beta$.

When the frequency becomes larger than $2\omega_F$, the real part of the conductivity approachs $\sigma_0$ while the
imaginary part vanishs. In this regime, the role of the graphene sheets consists, essentially, in absorbing light leading to a 
vanishing transmission after few stacks.

\section{Conclusions \label{ultima_sec}}

 In conclusion, we have investigated light propagation in
1D disordered superlattices composed of dielectric stacks and graphene
sheets in between. We introduced disorder either in the graphene material 
parameters (compositional disorder), such as the Fermi energy, or in the widths
of the dielectric stacks (structural disorder). For both cases we derived an analytical
expression for the localization length $\xi$ and compared the results with numerical 
calculations based on the transfer matrix method. A very good agreement between numerics 
and the analytical expression was found.  We demonstrated that, for structural disorder and
when the impedances of the layers are equal, the localization length does not follow the
well-known asymptotic behaviour $\xi \propto \omega^{-2}$. Rather, it exhibits an oscillatory
dependence on frequency, as a result of the presence of the Drude term in the graphene conductivity. 
Also in the impedance matching regime, we show that graphene has an important impact on the Brewster
modes, anomalously delocalised modes at given frequencies and  incident 
angles at which $\xi$ diverges.
Indeed, the presence of graphene induces additional reflections inside the disordered medium, leading
to a strong attenuation of the Brewster modes. We investigated how intra and interband transitions in 
the graphene conductivity impact on $\xi$, identifying the regimes where Anderson localization and 
absorption dominates light transmission. Altogether, our findings unveil the role of graphene on 
Anderson localization of light, paving the way for the design of graphene-based, disordered 
photonic devices in the THz spectral range.





\section*{Acknowledgements}

We thank W. Kort-Kamp for useful discussions. A. J. Chaves acknowledge the scholarship from the Brazilian
agency  CNPq (Conselho Nacional de Desenvolvimento
Cient\'ifico e Tecnol\'ogico). N.M.R.P. acknowledges financial 
support from the Graphene Flagship Project (Contract No. CNECT-ICT-604391). F.A.P. thanks the Optoelectronics Research Centre and Centre for Photonic Metamaterials, University of Southampton, for the hospitality, and CAPES for funding his visit (Grant No. BEX 1497/14-6). F.A.P. also acknowledges CNPq (Grant No. 303286/2013-0) for financial support.

\begin{appendix}

\section{Matrix transformation} \label{matrix}

The relation $\psi^{n+1}=M^n\psi^n$ can be interpreted as a discrete set of points
in the phase space $\psi_R,\psi_L$. With the transformation $M_\text{real}$:
\be
M_\text{real}=\frac{1}{2}\begin{pmatrix} 1-i &&1+i \\ -1+i && 1+i\end{pmatrix},
\ee
the matrix $M_\text{real} M^n M_\text{real}^{-1}$ is now real, and defining
${\psi^\prime}^n=M_\text{real} \psi^n$, we have in the phase space
$\psi_R^\prime,\psi_L^\prime$ that in the system without disorder
the trajectory is given by a ellipse. From this we can find a transformation
$M_\text{circle}$ to a circle:
\be
M_\text{circle}=\begin{pmatrix} v^{-1} \cos\tau && v\sin\tau\\-v^{-1}\sin\tau &&v\cos\tau\end{pmatrix},
\ee
where:
\bea
v^2=-\frac{\sin\gamma}{\cosh\phi^0_1\sin\phi_2^0+\sinh\phi^0_1},\nonumber\\
\tau=\frac{\pi}{4}-\frac{\phi_3^0}{2},
\eea
and making $[Q \,\,\,P]^T=M_\text{circle}\psi^\prime $,

\begin{flalign}
\begin{pmatrix} Q_n\\P_n\end{pmatrix}=\begin{pmatrix} v^{-1} \cos\tau && v\sin\tau\\-v^{-1}\sin\tau &&v\cos\tau\end{pmatrix}\begin{pmatrix} x_n\\y_n\end{pmatrix},
\end{flalign}

\section{Lyapunov Exponent} \label{formalism}

The Lyapunov exponent is given by:
\be
\lambda=\frac{1}{2}\left\langle Y_1+Y_2\cos2\Theta_n+Y_3\sin2\Theta_n-\frac{1}{4}Y_2^2-\frac{1}{4}Y_3^2 \right\rangle,
\ee
where
\be
Y_1=\frac{1}{\sin^2\gamma}\left[U_1\delta\phi_1^2+U_2\delta\phi_2^2+U_3\delta\phi_3^2+U_4\delta\phi_1\delta\phi_2\right], \label{y1}
\ee
with:
\bea
U_1&=&2\sin^2\phi_2^0,\\
U_2&=&2\sinh^2\phi_1^0\cos^2\gamma, \\
U_3&=&2\sinh^2\phi^0_1\sin^2\gamma,\\
U_4&=&-\sinh2\phi^0_1\sin2\phi_2^0,
\eea
\be
Y_2=\left[-2\sin\phi_2^0\delta\phi_1+\cos\phi_2^0\sinh2\phi_1^0\left(\delta\phi_2-\delta\phi_3\right)\right],\label{y2}
\ee
\be
Y_3=2\frac{\sinh\phi^0_1\left(\cos^2\gamma\delta\phi_2 +\sin^2\gamma\delta\phi_3\right)-\cos\gamma\sin\phi_2^0\delta\phi_1}{-\sin\gamma}.\label{y3}
\ee
the angle $\Theta$ obeys the recurrence equation:
\be
\Theta_{n+1}=\Theta_n-\gamma+\epsilon_n\csc\gamma, \label{eqTheta}
\ee
with:
\bea
\epsilon_n=&\left[\cos\gamma\sinh\phi^0_1\delta\phi_2-\sin\phi_2^0\delta\phi_1\right]\cos\left(2\Theta_n-\gamma\right)+\nonumber\\&\sinh\phi^0_1\cos\gamma\sin(2\Theta_n-\gamma)\delta \phi_3,
\eea

\section{Photonic Crystal} \label{phocry}
\subsection{Unit cell made of two different dielectrics and a graphene sheet at the interfaces}

The transfer matrix 
 whose elements are \cite{Zhan2013}
\footnote{there are some typos in the transfer matrix elements given in reference  \cite{Zhan2013}} \:
\bea
m_{11}^j&=&\left[A^x_-\cos\alpha_2+i(\chi+C^x_+)\sin\alpha_2\right]e^{-i\alpha_1},  \nonumber\\
m_{12}^j&=&\left[B^x\cos\alpha_2+i\left(\Delta+D^x\right)\sin\alpha_2\right]e^{i\alpha_1}, \nonumber \\
m_{21}^j&=&\left[-B^x\cos\alpha_2-i\left(\Delta+D^x\right)\sin\alpha_2\right]e^{-i\alpha_1},  \nonumber\\
m_{22}^j&=&\left[A^x_+\cos\alpha_2-i\left(\chi+C_-^x\right)\sin\alpha_2\right]e^{i\alpha_1}, \label{mtransfer}
\label{mFull}
\eea
where $x=\text{TE},\text{TM}$ and the diverse parameters are given in appendix \ref{phocry}.

The Snell-Decartes law hold:
\be
\sqrt{\varepsilon_1\mu_1}\sin\theta_1=\sqrt{\varepsilon_2\mu_2}\sin\theta_2, \label{snell}
\ee
and the dispersion relation is given by:
\bea
\cos\gamma=\cos\alpha_1\cos\alpha_2-\left(\chi+2f^2\beta_1^x\beta_2^x\right)\sin\alpha_1\sin\alpha_2\nonumber\\
+2if\left(\beta_1^x\cos\alpha_1\sin\alpha_2+\beta_2^x\cos\alpha_2\sin\alpha_1\right). \label{disrel}
\eea

\be
\beta_i^{TM}=Z_i^{TM}\,, \,\,\, \beta_i^{TE}=\frac{1}{Z_i^{TE}}, \label{betas}
\ee
where:
\bea
k_i&=&\sqrt{\varepsilon_i\mu_i}\omega/c\cos\theta_i,\nonumber\\
\alpha_i&=&k_i z_i,\nonumber\\
A^x_\pm&=&(1\pm2f\beta^x_1)\nonumber,\\
B^x&=&2f\lambda^x \beta_1^x, \nonumber\\
C^x_\pm&=&\pm2f\beta^x_2+2f^2 \beta^x_1 \beta^x_2,\nonumber\\
D^x&=&2f^2 \lambda^x\beta_1^x\beta_2^x,\nonumber\\
\eta^x&=&\frac{Z_1^x}{Z_2^x},\nonumber\\
\Delta^x&=&\frac{1}{2}\left(\eta^x-{\eta^x}^{-1}\right),\nonumber\\
\chi^x&=&\frac{1}{2}\left(\eta^x+{\eta^x}^{-1}\right),\nonumber\\
f&=&\frac{\sigma c\mu_0}{2}, \nonumber\\
Z^\text{TE}_i&=&\frac{\sqrt{\mu_i\varepsilon_i}}{\mu_i}\cos\theta_i, \nonumber\\
Z^\text{TM}_i&=&\frac{\sqrt{\mu_i\varepsilon_i}}{\varepsilon_i}\cos\theta_i,
\eea
with $\lambda^\text{TM}=+1,\lambda^\text{TE}=-1$.

\subsection{Unit cell made of one dielectric and a graphene sheet at the interface}

When there is only one dielectric, with width $z$ and $\varepsilon,\mu$ permissivity and permeability, intercalated by graphene sheets, the
 transfer matrix is given by:
\be
M=\begin{pmatrix} (1-\beta^x f)e^{i\alpha} && -\lambda^x \beta^x f e^{i\alpha} \\
                          \lambda^x \beta^x f e^{-i\alpha} && (1+\beta^x f)e^{-i\alpha}\end{pmatrix}, \label{1dg}
\ee
where $\alpha=\sqrt{\mu\varepsilon} z\cos\theta$
with the dispersion relation:
\be
\cos\gamma=\cos\alpha-i\beta^x f\sin\alpha. \label{cosg1d}
\ee

\section{Graphene Optical Conductivity} \label{grcond}

For completeness we give here the expressions for the optical conductivity of graphene,
whose derivation
can be found elsewhere \cite{Peres2010,Bludov20132}. The graphene optical conductivity of graphene
is a sum of a Drude term, $\sigma_D$, and an inter-band contribution, $\sigma_I$, reading:
\be
\sigma=\sigma_D+\sigma_I, \label{gcond}
\ee
where the Drude term is given by:
\be
\frac{\sigma_D}{\sigma_0}= \frac{4\omega_F}{\pi}\frac{1}{\Gamma-i\omega}, \label{drudeg}
\ee
and the interband term $\sigma_I=\sigma_I^\prime+i\sigma^{\prime\prime}_I$
have the real part
\be
\frac{\sigma_I^\prime}{\sigma_0}=\left(1+\frac{1}{\pi}\arctan\frac{\omega-2\omega_F}{\Gamma}-\frac{1}{\pi}\arctan\frac{\omega+2\omega_F}{\Gamma}\right),
\ee
and the imaginary part
\be
\frac{\sigma_I^{\prime\prime}}{\sigma_0}=-\frac{1}{2\pi}\ln\frac{(2\omega_F+\omega)^2+\Gamma^2}{(2\omega_F-\omega)^2+\Gamma^2}, \label{interimag}
\ee
with the Fermi frequency given by
\be
\omega_F=\frac{|E_F|}{\hbar}.
\ee

\end{appendix}



%

\end{document}